\documentclass[%
 aip,jcp,
 amsmath,amssymb,
 reprint,%
]{revtex4-1}

\usepackage{graphicx}
\usepackage{dcolumn}
\usepackage{bm}

\usepackage[utf8]{inputenc}
\usepackage[T1]{fontenc}
\usepackage{mathptmx}

\usepackage{mathtools}
\usepackage{relsize}
\DeclarePairedDelimiter\bra{\langle}{\rvert}
\DeclarePairedDelimiter\ket{\lvert}{\rangle}
\DeclarePairedDelimiterX\braket[2]{\langle}{\rangle_N}{#1 \delimsize\vert #2}
\newcommand{\ham}[1]{{\hat H}_{#1}}
\usepackage{multirow}
\begin{document}

\preprint{AIP/123-QED}

\title{Polaritonic normal modes in Transition State Theory}

\author{J. A. Campos-Gonzalez-Angulo}
 \affiliation{Department of Chemistry and Biochemistry. University of California San Diego. La Jolla, California 92093, USA}
\author{J. Yuen-Zhou}
 \email{joelyuen@ucsd.edu}
 \homepage{http://yuenzhougroup.ucsd.edu}
\affiliation{Department of Chemistry and Biochemistry. University of California San Diego. La Jolla, California 92093, USA}%


\begin{abstract}
A series of experiments demonstrate that strong light-matter coupling between vibrational excitations in isotropic solutions of molecules and resonant infrared optical microcavity modes leads to modified thermally-activated kinetics. However, Feist and coworkers [\emph{Phys. Rev. X.}, \textbf{9}, 021057(2019)] have recently demonstrated that, within transition state theory, effects of strong light-matter coupling with reactive modes are  electrostatic, and essentially independent of light-matter resonance or even of the formation of vibrational polaritons. To analyze this puzzling theoretical result in further detail, we revisit it under a new light, invoking a normal mode analysis of the transition state and reactant configurations for an ensemble of an arbitrary number of molecules in a cavity, obtaining simple analytical expressions that produce similar conclusions as Feist. While these effects become relevant in optical microcavities if the molecular dipoles are anisotropically aligned, or in cavities with extreme confinement of the photon modes, they become negligible for isotropic solutions in microcavities. It is concluded that further studies are necessary to track the origin of the experimentally observed kinetics.

\end{abstract}

\maketitle


\section{\label{sec:intro}Introduction}

Multiple experimental results show that reactions taking place inside of optical microcavities proceed with different kinetics than outside of them.\cite{Thomas2016,Hiura2018,Lather2019,Thomas2019,Vergauwe2019,Hirai2020} Rate modification seems to require that the confined electromagnetic mode couples to one of the varieties of molecular vibrational modes present in the reactive medium.\cite{Thomas2019} For reactions in solution, where molecules are isotropically distributed, this coupling is maximized under resonant conditions, \textit{i.e.}, when the cavity is tuned to a vibrational frequency in the molecules. Also, the effect on the kinetics has been observed to increase as the collective coupling intensifies, as a consequence of the large number of molecules present in a sample.\cite{Thomas2016}
These observations are reminiscent of the description of light-matter coupling in terms of hybrid states known as polaritons,\cite{Ebbesen2016,Ribeiro2018,Feist2018,Flick2018,Ruggenthaler2018,Herrera2020} which successfully explains the optical properties of these systems.\cite{Shalabney2015,Casey2016,Dunkelberger2016,Xiang2018,Erwin2019} Recently, it has been suggested that a class of nonadiabatic charge transfer reactions would experience a catalytic effect from resonant collective coupling between high-frequency modes and infrared cavity modes; the mechanism relies on the formation of vibrational polaritons which feature reduced activation energies compared to the bare molecules.\cite{CamposGonzalezAngulo2019,Phuc2019}

However, a large class of reactions fall in the adiabatic regime, where the potential energy surfaces of the electronic ground and excited states are well-separated. These reactions should be accurately described by a transition state theory (TST)\cite{Truhlar1996,Nitzan2006,Vaillant2019} that accounts for vibrational strong coupling (VSC). Feist and coworkers have in fact developed a theoretical framework with the essential ingredients to capture the action of a confined electromagnetic field on chemical processes such as nucleophyllic substitution.\cite{Galego2019,Climent2019} Within this framework, they find that the presence of a cavity mode modifies the reactive potential energy surface, thus predicting conditions for increase and decrease of reaction rates. However, according to their results, resonance is not essential for this modification to take place. Furthermore, the effect depends on the intensity of the single-molecule coupling, and cooperativity can only occur under conditions such as the anisotropic alignment of the permanent dipoles, an unlikely condition for the aforementioned reported experiments.\cite{Li2020} Remarkably, Feist's formalism excludes the language of polaritons. In fact, they concede that polaritonic degrees of freedom appear inconsequentially in the form of normal modes near the equilibrium configurations of the system, and that the effects are of the (Casimir-Polder) electrostatic type.\cite{Galego2019} In the present work, we restate their formalism bringing the polaritonic modes into the limelight; we take advantage of the polaritonic framework to expand the formalism and obtain simple and physically intuitive analytical TST expressions that describe the modified collisional prefactors and activation energies in terms of light and matter parameters. Our results are in line with the predictions of \cite{Galego2019,Climent2019}, highlighting that further work must be carried out to understand the difference between experiment and theory in the context of thermally-activated reactions under VSC.
\section{\label{sec:theo}Theory}

\begin{figure}
\includegraphics[width=\linewidth]{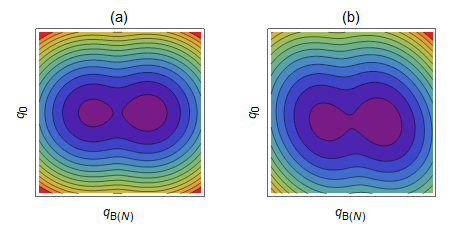}
\caption{Effect of VSC on a reactive potential energy surface. a) Asymmetric double well potential uncoupled to an orthogonal harmonic cavity mode. b) Same as in (a) but with non-zero light-matter coupling. The distortion of the wells reveals the redefinition of normal modes from cavity and molecule to upper and lower polaritons.\label{fig:pes}}
\end{figure}

According to TST, the rate constant at temperature $T$ is defined as\cite{Wigner1938,Haenggi1990,Pollak2005,Arnaut2006,Henriksen2018}
\begin{equation}\label{eq:ktst}
k_\textrm{TST}=\frac{k_B T}{2\pi \hbar}\frac{Z_\ddag}{Z_\textrm{eq}}\textrm{e}^{-\frac{E_a}{k_B T}},
\end{equation}
where $k_B$ and $\hbar$ are the Boltzmann and reduced Planck constants, respectively. $Z_\ddag$ is the partition function of the transition state (TS) without the contribution of the reactive mode, and $Z_\textrm{eq}$ is the total partition function of the reactant state. $E_a=V_\ddag+\frac{1}{2}\sum_i\hbar\omega_{i,\ddag}-V_\textrm{eq}-\frac{1}{2}\sum_j\hbar\omega_{j,\textrm{eq}}$ is the activation energy, where the frequency $\omega_{i,r}$ corresponds to the square root of the $i$-th positive eigenvalue of the Hessian of the potential energy surface evaluated at the state $r$.
We will determine how the rate constant changes for a thermally-activated process in which the reactant is a heteronuclear diatomic molecule, when it takes place inside an optical microcavity. While the following analysis can be straightforwardly generalized for a multimode system, we will treat only the simplest case for the sake of conceptual clarity. Such a system with $N$ identical reactant molecules can be described by the Hamiltonian\cite{Flick2017,Galego2019}
\begin{equation}\label{eq:ham}
\ham{}=\ham{\textrm{EM}}+\sum_{i=1}^N\left(\ham{\textrm{mol}}^{(i)}+\hat{V}_\textrm{int}^{(i)}\right),
\end{equation}
 where $\ham{\textrm{EM}}=\hbar\omega_0\left(\hat a_0^\dag \hat a_0+\tfrac{1}{2}\right)$ characterizes a confined electromagnetic field of frequency $\omega_0$, and creation and annihilation operators $\hat{a}_0^\dag$ and $\hat{a}_0$, respectively. $\ham{\textrm{mol}}^{(i)}=\hat{T}_\textrm{nuc}^{(i)}+\hat{V}_\textrm{nuc}^{(i)}+\hat{T}_\textrm{elec}^{(i)}+\hat{V}_\textrm{elec}^{(i)}+\hat{V}_\textrm{nuc-elec}^{(i)}$ is the Hamiltonian of the $i$-th molecule containing the kinetic, $\hat{T}$, and potential, $\hat{V}$, energies of the nuclear and electronic degrees of freedom, as well as their Coulomb interaction. The coupling between light and matter is given by $\hat V_\textrm{int}^{(i)}=g\omega_0 \hat q_0\bm{\epsilon}\cdot\bm{\hat \mu}_i$, where $\hat q_0=\sqrt{\frac{\hbar}{2\omega_0}}\left(\hat{a}_0^\dag+\hat{a}_0\right)$, and $g=-(\mathcal{V}\varepsilon_0)^{-1/2}$ is the coupling constant, with $\mathcal{V}$ the mode volume and $\varepsilon_0$ the vacuum permittivity; $\bm{\epsilon}$ is the polarization vector of the cavity field, and $\bm{\hat{\mu}}_i$ is the molecular vibrational electric dipole moment.
In the (cavity) Born-Oppenheimer approximation,\cite{Flick2017a,Ruggenthaler2018} the ground state potential energy for the electronic Schrödinger equation with Hamiltonian $\hat{H}_\textrm{elec}=\hat{H}-\sum_{i=1}^N\hat{T}_\textrm{nuc}$, can be parameterized in terms of the nuclear coordinates, $\mathbf{R}$, and the photon coordinate $q_0$, which is an eigenvalue of the operator $\hat q_0$. Thus, the potential energy surface governing the nuclear degrees of freedom (Fig. \ref{fig:pes}) becomes
\begin{equation}\label{eq:pes}
V(\mathbf{R},q_0)=\sum_{i=1}^N V_\textrm{nuc}(\mathbf{R}_i)+\frac{\omega_0^2}{2}q_0^2+\omega_0 g q_0 \bm{\epsilon}\cdot\sum_{i=1}^N\bm{\mu}(\mathbf{R}_i).
\end{equation}
In writing Eqs. \eqref{eq:ham} and \eqref{eq:pes} we have neglected the diamagnetic term arising from the Power-Zienau-Woolley transformation.\cite{CohenTannoudji1989} Its relevance for problems in the current context is explored in detail in Refs. \onlinecite{Schaefer2019,Li2020}. Nevertheless, since even in the ultrastrong regime, light-matter coupling \emph{per molecule} is much smaller than the vibrational transition energies,\cite{MartinezMartinez2018} the inclusion of such term should only account for slight modifications to the formalism that leave the findings unchanged.

In the neighborhood of the equilibrium configuration of the reactants, $\mathbf{R}_\textrm{eq}$, the potential is reasonably well described by a second order expansion while the dipole moment can be approximated to first order:
\begin{equation}\label{eq:veq}
\begin{split}
V(\mathbf{R}\approx\mathbf{R}_\textrm{eq},q_0)=&\sum_{i=1}^NV_\textrm{nuc}(\mathbf{R}_{i,\textrm{eq}})+\frac{\omega_\textrm{eq}^2}{2}\sum_{i=1}^N q_i^2+\frac{\omega_0^2}{2}q_0^2\\
&+\omega_0 g q_0\sum_{i=1}^N\left(\mu_{i,\textrm{eq}}+\mu_{i,\textrm{eq}}' q_i\right),
\end{split}
\end{equation}
where $q_i$ is the mass-reduced bond elongation with respect to the equilibrium length of the $i$-th molecule, $\omega_\textrm{eq}^2=\left.\frac{\partial^2V_\textrm{nuc}^{(i)}}{\partial q_i^2}\right\rvert_0$, $\mu_{i,\textrm{eq}}=\bm{\epsilon}\cdot\bm{\mu}(\mathbf{R}_{i,\textrm{eq}})$, and $\mu'_{i,\textrm{eq}}=\bm{\epsilon}\cdot\left.\frac{\partial\bm{\mu}(\mathbf{R}_i)}{\partial q_i}\right\rvert_0$.
We note that this expansion excludes the polarizability term present in the perturbative treatment by \citet{Galego2019}; however, as we shall see, this omission does not affect the main conclusions.

Differentiation of Eq. \eqref{eq:veq} yields
\begin{subequations}
\begin{align}
\frac{\partial V}{\partial q_0}&=\omega_0^2 q_0+\omega_0 g\sum_{i=1}^N\left(\mu_{i,\textrm{eq}}+\mu_{i,\textrm{eq}}' q_i\right)\\
\frac{\partial V}{\partial q_j}&=\omega_\textrm{eq}^2 q_j+\omega_0 g q_0\mu_{j,\textrm{eq}}'\quad 1\leq j\leq N;
\end{align}
\end{subequations}
therefore, at the new minimum, $\mathbf{R}^\textrm{VSC}_\textrm{eq}$, close to $\mathbf{R}_\textrm{eq}$, the coordinates fulfill
\begin{multline}\label{eq:mat1}
\begin{pmatrix}
\omega_0^2&\omega_0 g\sqrt{N\left\langle\mu_\textrm{eq}^{\prime2}\right\rangle_N}\\
\omega_0 g\sqrt{N\left\langle\mu_\textrm{eq}^{\prime2}\right\rangle_N}&\omega_\textrm{eq}^2
\end{pmatrix}
\begin{pmatrix}
q_0\\q_{\textrm{B}(N)}
\end{pmatrix}\\
=-\omega_0 g N\left\langle\mu_\textrm{eq}\right\rangle_N
\begin{pmatrix}
1\\0
\end{pmatrix},
\end{multline}
where $\langle x\rangle_N=\frac{1}{N}\sum_{i=1}^N x_i$, and the bright molecular mode is given by $q_{\textrm{B}(N)}=\sqrt{\frac{N}{\left\langle\mu_\textrm{eq}^{\prime2}\right\rangle_N}}\left\langle\mu'_\textrm{eq}q\right\rangle_N$.

The coefficient matrix in Eq. \eqref{eq:mat1} corresponds to the Hopfield-Bogoliubov form of the Dicke model in the normal phase\cite{Emary2003,BastarracheaMagnani2014}; therefore, its diagonalization gives rise to polariton modes, as shown in Fig. \ref{fig:pes}. To be specific, Eq. \eqref{eq:mat1} can be rewritten as
\begin{equation}
\begin{pmatrix}
\omega_{+(N)}^2&0\\
0&\omega_{-(N)}^2
\end{pmatrix}
\begin{pmatrix}
q_{+(N)}\\q_{-(N)}
\end{pmatrix}
=-\omega_0 g N\left\langle\mu_\textrm{eq}\right\rangle_N
\begin{pmatrix}
\cos\theta_N\\ \sin\theta_N
\end{pmatrix},
\end{equation}
where $\omega_{\pm(N)}^2=\frac{1}{2}\left[\omega_0^2+\omega_\textrm{eq}^2\pm\sqrt{4\omega_0^2 g^2 N\left\langle\mu_\textrm{eq}^{\prime2}\right\rangle_N+\left(\omega_0^2-\omega_\textrm{eq}^2\right)^2}\right]$ is the frequency squared of the upper(lower) polaritonic mode,
$\begin{pmatrix}q_{+(N)}\\q_{-(N)}\end{pmatrix}=\begin{pmatrix}\cos\theta_N&-\sin\theta_N\\ \sin\theta_N&\cos\theta_N\end{pmatrix}\begin{pmatrix}q_0\\q_{\textrm{B}(N)}\end{pmatrix}$ are the polaritonic mode coordinates, and $\theta_N=-\frac{1}{2}\arctan\frac{2\omega_0 g\sqrt{N\left\langle\mu_\textrm{eq}^{\prime2}\right\rangle_N}}{\omega_0^2-\omega_\textrm{eq}^2}$ is the mixing angle.

Equation \eqref{eq:veq} can be recast using this new set of coordinates in the form
\begin{multline}\label{eq:Vpol}
V(\mathbf{R}\approx\mathbf{R}_\textrm{eq},q_0)=\sum_{i=1}^NV_\textrm{nuc}(\mathbf{R}_{i,\textrm{eq}})+\frac{\omega_\textrm{eq}^2}{2}\sum_{k=1}^{N-1} q_{\textrm{D}(N)}^{(k)2}\\
+\frac{\omega_{+(N)}^2}{2}q_{+(N)}^2+\frac{\omega_{-(N)}^2}{2}q_{-(N)}^2\\
+\omega_0 g N\left\langle\mu_\textrm{eq}\right\rangle_N\left(\cos\theta_N q_{+(N)}+\sin\theta_N q_{-(N)}\right),
\end{multline}
where $q_{\textrm{D}(N)}^{(k)}=\sum_{i=1}^N c_{ki}q_i$ are the dark vibrational modes, with the coefficients $c_{ki}$ fulfilling $\sum_{i=1}^N\mu^{\prime*}_{i,\textrm{eq}}c_{ki}=0$ and $\sum_{i=1}^N c_{k'i}^* c_{ki}=\delta_{k'k}$.
Evaluating the potential in Eq. \eqref{eq:Vpol} at $\mathbf{R}^\textrm{VSC}_\textrm{eq}$ yields
\begin{equation}
V_\textrm{eq}^\textrm{VSC}=\sum_{i=1}^NV_\textrm{nuc}(\mathbf{R}_{i,\textrm{eq}})-\left(\frac{\omega_0\omega_\textrm{eq}}{\omega_{+(N)}\omega_{-(N)}}gN\langle\mu_\textrm{eq}\rangle_N\right)^2.
\end{equation}
We note that the modification to the potential is proportional to the ratio of the determinants of the Hessian without and with light-matter coupling, which acts as a measure of the redefinition of the normal modes. Additionally, the presence of the permanent dipole reveals the electrostatic nature of this effect.
 
Without loss of generality, let us assume that the molecule with label $N$ undergoes a reaction. The potential energy surface in the neighborhood of the TS configuration, $\mathbf{R}_\ddag$, is
\begin{multline}\label{eq:vtst}
V(\mathbf{R}\approx\mathbf{R}_\ddag,q_0)=\sum_{i=2}^NV_\textrm{nuc}(\mathbf{R}_{i,\textrm{eq}})+V_\textrm{nuc}(\mathbf{R}_{N,\ddag})\\
+\frac{\omega_\textrm{eq}^2}{2}\sum_{i=1}^{N-1} q_i^2+\frac{\omega_0^2}{2}q_0^2+\frac{\omega_\ddag^2}{2}q_N^2\\
+\omega_0 g q_0 \left[\sum_{i=1}^{N-1}\left(\mu_{i,\textrm{eq}}+\mu'_{i,\textrm{eq}} q_i\right)+\mu_\ddag+\mu'_\ddag q_N\right].
\end{multline}
Here, $\omega_\ddag^2=\left.\frac{\partial^2V_\textrm{nuc}^{(N)}}{\partial q_N^2}\right\rvert_{q_\ddag}<0$ is the squared frequency of the unstable mode, $\mu_\ddag=\bm{\epsilon}\cdot\bm{\mu}(\mathbf{R}_{N,\ddag})$, and $\mu'_\ddag=\bm{\epsilon}\cdot\left.\frac{\partial\bm{\mu}(\mathbf{R}_N)}{\partial q_N}\right\rvert_{q_\ddag}$.

Applying the previous treatment to the potential energy surface in the saddle point, $\mathbf{R}^\textrm{VSC}_\ddag$, the coordinates fulfill
\begin{widetext}
\begin{equation}\label{eq:matst}
\begin{pmatrix}
\omega_0^2&\omega_0 g\sqrt{(N-1)\left\langle\mu_\textrm{eq}^{\prime2}\right\rangle_{N-1}}&\omega_0 g\mu'_\ddag\\
\omega_0 g\sqrt{(N-1)\left\langle\mu_\textrm{eq}^{\prime2}\right\rangle_{N-1}}&\omega_\textrm{eq}^2&0\\
\omega_0 g\mu'_\ddag&0&\omega_\ddag^2
\end{pmatrix}
\begin{pmatrix}
q_0\\q_{\textrm{B}(N-1)}\\q_N
\end{pmatrix}
=-\omega_0 g\left[(N-1)\left\langle\mu_\textrm{eq}\right\rangle_{N-1}+\mu_\ddag\right]
\begin{pmatrix}
1\\0\\0
\end{pmatrix}.
\end{equation}
\end{widetext}

For typical values of the transition dipole moments, the off-diagonal terms that depend on $N$ remain significant since the number of molecules per cavity mode is estimated between $10^6$ and $10^{10}$.\cite{Pino2015,Daskalakis2017} The term $g\omega_0\mu'_\ddag$ is several orders of magnitude smaller, and we can neglect it to recover a polaritonic picture where
\begin{multline}\label{eq:mtstpol}
\begin{pmatrix}
\omega_{+(N-1)}^2&0&0\\
0&\omega_{-(N-1)}^2&0\\
0&0&\omega_\ddag^2
\end{pmatrix}
\begin{pmatrix}
q_{+(N-1)}\\q_{-(N-1)}\\q_N
\end{pmatrix}\\
\approx-\omega_0 g\left[(N-1)\left\langle\mu_\textrm{eq}\right\rangle_{N-1}+\mu_\ddag\right]
\begin{pmatrix}
\cos\theta_{N-1}\\ \sin\theta_{N-1}\\0
\end{pmatrix}
\end{multline}
at $\mathbf{R}^\textrm{VSC}_\ddag$. Thus, the potential at the saddlepoint becomes
\begin{equation}
\begin{split}
V_\ddag^\textrm{VSC}=&\sum_{i=1}^{N-1}V_\textrm{nuc}(\mathbf{R}_{i,\textrm{eq}})+V_\textrm{nuc}(\mathbf{R}_{N,\ddag})\\
&-\left(\frac{\omega_0\omega_\textrm{eq}}{\omega_{+(N-1)}\omega_{-(N-1)}}g\left[(N-1)\left\langle\mu_\textrm{eq}\right\rangle_{N-1}+\mu_\ddag\right]\right)^2.
\end{split}
\end{equation}

From Eqs. \eqref{eq:veq}, \eqref{eq:vtst} and \eqref{eq:mtstpol}, it follows that the step to the TS can be written as
\begin{multline}
\textrm{UP}_N+\textrm{LP}_N+\sum_{k=1}^{N-1} D_N^{(k)}\longrightarrow\\
\textrm{UP}_{N-1}+\textrm{LP}_{N-1}+\sum_{k'=1}^{N-2} D_{N-1}^{(k')}+R_N^\ddag
\end{multline}
where $R_N^\ddag$ represents the reactive molecule in the TS. Therefore, the rate constant should include the partition functions of the whole ensemble of molecules coupled to light; however, as we will see, since only one molecule undergoes the reaction, the ratio of partition functions simplifies to an intelligible expression in terms of the single molecule $k_\textrm{TST}$.

Outside of the cavity the rate constant takes the form
\begin{equation}\label{eq:ktst}
\begin{split}
k_\textrm{TST}=&\frac{k_B T}{\pi \hbar}\frac{Q_\ddag}{Q_\textrm{eq}}\sinh\left(\frac{\hbar\omega_\textrm{eq}}{2k_B T}\right)\\
&\times\exp\left(-\frac{V_\textrm{nuc}\left(\mathbf{R}_{N,\ddag}\right)-V_\textrm{nuc}\left(\mathbf{R}_{N,\textrm{eq}}\right)}{k_B T}\right),
\end{split}
\end{equation}
where the ratio $Q_\ddag/Q_\textrm{eq}$ captures all the information from the translational and rotational degrees of freedom (for a 1D system comprised of the reactive mode only, $Q_\ddag=Q_\textrm{eq}$). To characterize the effect of the cavity mode on the kinetics, we define
\begin{equation}
k_\textrm{TST}^\textrm{VSC}=\kappa_Nk_\textrm{TST},
\end{equation}
where the ratio of rate constants is given by
\begin{subequations}
\begin{equation}
\kappa_N=A_\textrm{VSC}(T)\exp\left(-\frac{\Delta V_\textrm{VSC}+\Delta E_0^\textrm{VSC}}{k_B T}\right),
\end{equation}
with prefactor
\begin{equation}
A_\textrm{VSC}(T)=\frac{\sinh\left(\hbar\omega_{+(N)}/2k_B T\right)\sinh\left(\hbar\omega_{-(N)}/2k_B T\right)}{\sinh\left(\hbar\omega_{+(N-1)}/2k_B T\right)\sinh\left(\hbar\omega_{-(N-1)}/2k_B T\right)},
\end{equation}
cavity-induced potential energy difference
\begin{multline}\label{eq:DVvsc}
\Delta V_\textrm{VSC}=\omega_0^2\omega_\textrm{eq}^2g^2\\
\times\left[\left(\frac{N\langle\mu_\textrm{eq}\rangle_N}{\omega_{+(N)}\omega_{-(N)}}\right)^2-\left(\frac{(N-1)\left\langle\mu_\textrm{eq}\right\rangle_{N-1}+\mu_\ddag}{\omega_{+(N-1)}\omega_{-(N-1)}}\right)^2\right],
\end{multline}
and zero-point-energy difference
\begin{equation}
\Delta E_0^\textrm{VSC}=\frac{\hbar\omega_{+(N-1)}+\hbar\omega_{-(N-1)}-\hbar\omega_{+(N)}-\hbar\omega_{-(N)}}{2}.
\end{equation}
\end{subequations}

\begin{figure}
\includegraphics[width=\linewidth]{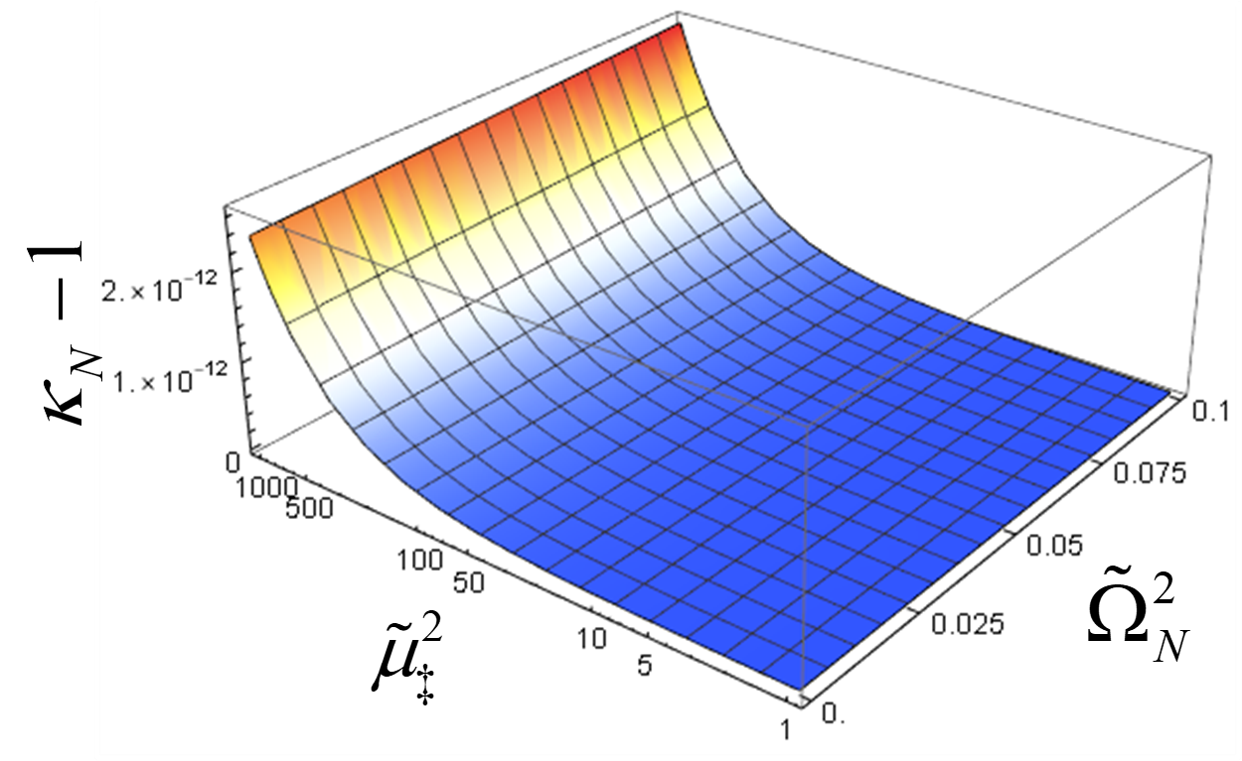}
\caption{Ratio of rate constants as a function of the permanent dipole in the TS, and the collective light-matter coupling. $\tilde{\mu}_\ddag=\mu_\ddag/\lvert\bra{n+1}\hat{\mu}_\textrm{eq}\ket{n}\rvert$ is the permanent dipole moment in the TS normalized with respect to the transition dipole moment in the equilibrium configuration, and $\tilde{\Omega}_N=g\sqrt{N\left\langle_\textrm{eq}^2\right\rangle}/\omega_\textrm{eq}$ is the light-matter coupling normalized with respect to the frequency in the same configuration. Over the span of the weak and strong coupling regimes, and a wide range of values of the TS dipole, the transmission coefficient remains close to 1. For this calculation, $\omega_{eq}=2000~\textrm{cm}^{-1}$, $N=10^9$, and $k_B T=208.5~\textrm{cm}^{-1}$.\label{fig:transcoefN}}
\end{figure}

As stated before, $N\gg1$. In this limit, 
$A_\textrm{VSC}(T)\approx1$, $\Delta E_0^\textrm{VSC}\approx0$, and 
the ratio of rate constants becomes
\begin{equation}\label{eq:transcoef}
\kappa_N\approx\exp\left[\frac{\left(\omega_\textrm{eq}g\mu_\ddag\right)^2}{\left(\omega_\textrm{eq}^2-g^2 N\left\langle\mu_\textrm{eq}^{\prime2}\right\rangle\right)k_B T}\right],
\end{equation}
where we have considered that, for typical reactions in liquid solution, the molecular dipoles are isotropically distributed; therefore, $\langle\mu_\textrm{eq}\rangle_N=0$. Regarding collective effects, in Fig. \ref{fig:transcoefN}, we show the ratio of rate constants as a function of the collective coupling and the permanent dipole moment of the TS. We can see that the variation of $\kappa_N$ throughout the span of the weak and strong light-matter coupling regimes is negligible. Furthermore, even over a huge range of possible values of $\mu_\ddag$, the ratio of rate constants remains too close to 1 to imply any observable change in the reaction rate. In contrast, note that in a sample with perfectly aligned dipoles, $\left\langle\mu_\textrm{eq}\right\rangle_N\neq0$, leading to substantial collective $O(N)$ contributions to $\Delta V_\textrm{VSC}$ [see Eq. \eqref{eq:DVvsc}]. Furthermore, regardless of dipole alignment, it can be shown that $\Delta V_\textrm{VSC}$ is independent of $\omega_0$, and is therefore unable to describe a resonant effect. 

From the previous analysis we reach the same conclusions of \citet{Galego2019}: effects of resonance between the cavity and the vibrational modes cannot be captured in a description at the level of TST, and the isotropic distribution of the permanent dipole moments negates the possibility of cooperative light-matter coupling effects. These results contrast with the situation of thermally-activated nonadiabatic charge transfer reactions, where the role of collective light-matter resonance in isotropic media is more evident. While we agree that the role of the polaritonic picture in our present analysis is rather shallow, it undoubtedly simplifies and clarifies the theoretical analysis. In conclusion, our results restate that a TST that takes into account strong coupling of the reactive mode to a resonant optical cavity mode is still insufficient to explain the experimental results  involving thermally-activated adiabatic reactions in Refs. \onlinecite{Thomas2016,Hiura2018,Lather2019,Thomas2019,Vergauwe2019,Hirai2020}.    

\begin{acknowledgments}
The authors thank Raphael F. Ribeiro, Luis A. Martínez-Martínez and Matthew Du for their insightful comments and discussions. This work was partially supported by the Defense Advanced Research Projects Agency under Award No. D19AC00011. JACGA also acknowledges support from UC-MEXUS/CONACYT through scholarship ref. 235273/472318. 
\end{acknowledgments}
\section*{Data Availability Statement}
Data sharing is not applicable to this article as no new data were created or analyzed in this study.
\appendix
\section{Single-molecule case}
When there is a single molecule per cavity mode, the only surviving coupling in Eq. \eqref{eq:matst} is that between the TS and the photon. In this case, the saddlepoint condition can be recast in terms of the eigenmodes as
\begin{equation}\label{eq:matst1}
\begin{pmatrix}
\omega_{-\ddag}^2&0\\
0&\omega_{+\ddag}^2
\end{pmatrix}
\begin{pmatrix}
q_{+\ddag}\\q_{-\ddag}
\end{pmatrix}
=-\omega_0 g\mu_\ddag
\begin{pmatrix}
\cos\theta_\ddag\\ \sin\theta_\ddag
\end{pmatrix},
\end{equation}
where $\omega_{-\ddag}^2<0<\omega_{+\ddag}^2$.
The potential energy evaluated at $\mathbf{R}^\textrm{VSC}_\ddag$ is
\begin{equation}
V_\ddag^\textrm{VSC}=V_\textrm{nuc}(\mathbf{R}_{\ddag})-\left(\frac{\omega_0\omega_\ddag}{\omega_{+\ddag}\omega_{-\ddag}}g\mu_\ddag\right)^2,
\end{equation}
which produces
\begin{subequations}
\begin{align}
A_\textrm{VSC}&=\frac{\sinh\left(\hbar\omega_{+}/2k_B T\right)\sinh\left(\hbar\omega_{-}/2k_B T\right)}{\sinh\left(\hbar\omega_{+\ddag}/2k_B T\right)\sinh\left(\hbar\omega_\textrm{eq}/2k_B T\right)},\\
\Delta V_\textrm{VSC}&=g^2\omega_0^2\left[\left(\frac{\omega_\textrm{eq}\mu_\textrm{eq}}{\omega_{+}\omega_{-}}\right)^2-\left(\frac{\omega_\ddag\mu_\ddag}{\omega_{+\ddag}\omega_{-\ddag}}\right)^2\right],\\
\Delta E_0^\textrm{VSC}&=\frac{\hbar\omega_{+\ddag}-\hbar\omega_{+}-\hbar\omega_{-}+\hbar\omega_\textrm{eq}}{2}.
\end{align}
\end{subequations}
It is worth noting that, despite $A_\textrm{VSC}(T)$ and $\Delta E_0^\textrm{VSC}$ deviating from 1 and 0, respectively, in the single-molecule limit, the effect is still off-resonant, thus reinforcing the findings in \citet{Galego2019}. In any case, the mode volumes and transition dipole moments required to modify a reaction rate are unrealistic unless we consider nano- and picocavities.

\bibliography{RefscTST}

\end{document}